# On intrinsic Stokes shift in wide GaN/AlGaN polar quantum wells

M. Jarema,[1,*] M. Gladysiewicz,[1] E. Zdanowicz,[1] E. Bellet-Amalric,[2] E. Monroy,[2] and R. Kudrawiec[1,**]

[1] *Faculty of Fundamental Problems of Technology, Wrocław University of Science and Technology, Wybrzeże Wyspiańskiego 27, 50-370 Wrocław, Poland*

[2] *Univ. Grenoble-Alpes, CEA, IRIG-DEPHY-PHELIQS-NPSC, 17 av. des Martyrs, 38054 Grenoble cedex 9, France*

*E-mail: michal.jarema@pwr.edu.pl

**E-mail: robert.kudrawiec@pwr.edu.pl

The interpretation of electromodulated reflectance (ER) spectra of polar quantum wells (QWs) is difficult even for homogeneous structures because of the built-in electric field. In this work we compare the room-temperature contactless ER and photoluminescence (PL) spectra of polar GaN/AlGaN QWs with the effective-mass band structure calculations. We show that the emission from the ground state transition is observed in PL but the ER is dominated by transitions between excited states. This effect results from the polarization-induced built-in electric field in QW that breaks the selection rules that apply to square-like QWs, allowing many optical transitions which cannot be separately distinguished in the ER spectrum. We develop the guidelines for the identification of optical transitions observed in PL and ER spectra. We conclude that an intrinsic Stokes shift, i.e., a shift between emission and absorption, is present even for homogeneous GaN/AlGaN QWs with large width, where the electron-hole wavefunction overlap for the fundamental transition is weak.

**I. INTRODUCTION**

Electromodulated reflectance (ER) spectroscopy (photoreflectance (PR) and contactless electroreflectance (CER)) is a very useful tool for characterization of semiconductor quantum wells (QWs).[1–11] Unlike photoluminescence (PL), ER is an absorption-like experiment that is insensitive to carrier localization. Therefore, comparing ER and PL spectra allows the determination of the Stokes shift energy, which can provide important information about the sample quality.[12,13] Moreover, ER probes the transitions between both ground and excited states, whereas ground transitions are dominant in low-temperature PL studies. The precise measurements of electronic transitions in QWs compared to theoretical calculations of the electronic structure allow the determination of relevant material parameters such as the band gap discontinuity.[8,14]

Polar GaN/AlGaN QWs are important due to their applications e.g. in interband devices like ultraviolet light-emitting diodes,[15–18] or intersubband devices operating in the infrared range.[19–22] Such QWs have been already investigated using ER techniques.[10,23–26] Unfortunately, the interpretation of ER spectra is still unclear or incomplete. Very often only the Franz–Keldysh oscillation[27] is analyzed to determine the built-in electric field in AlGaN barriers, but the portion of spectrum related to QW transitions is not interpreted, or it is unclear how the fitted ER resonances should be attributed to QW transitions.

The case of polar QWs is challenging because of the existence of a built-in electric field, which breaks the selection rules that apply to square QWs and allows additional interband transitions. Such transitions can be closer in energy than their thermal broadening ($\sim k_B T$), hence their resonances can overlap/merge in the ER spectra. In wide (< 2 nm) polar QWs, the internal electric field may also cause a significant separation of the ground electron and hole wavefunctions, which results in a small oscillator strength of the ground state (GS) transition. This can lead to a situation where the GS transition is very weak or almost not resolved in absorption-like spectra including ER spectra.

The aim of this paper is to provide a detailed interpretation of the ER spectra measured for polar QWs in the energy range of QW interband transitions. It will help to avoid a confusion associated with the identification of ER signal related to QW transitions and the determination/interpretation of the Stokes shift. It is worth noting that in the case of GaN/AlGaN QWs the interpretation is not trivial, although it is simpler than in case of InGaN/GaN QWs. In the case of wide GaN/AlGaN QWs, complications arising from inhomogeneous broadening[28] are minimized since the alloy fluctuations are eliminated in the active QW layer (GaN is a binary material), and the effect of 1-2 monolayer (ML) fluctuations of QW width can be neglected at the first approximation for >10 ML (≈ 2.5 nm) wide QWs.

This paper is structured as follows. After this brief introduction (section I), section II describes (A) the QW samples, their growth and characterization methods, and (B) the electronic structure calculations and the simulation of ER and PL spectra. In Section III, the measured spectra are analyzed and interpreted based on electronic band structure calculations. Next, we show that the CER spectrum can be reproduced/simulated using the calculation results (transition energies, strengths, broadenings) and fitting unknown phase factors. Additionally, the merging of resonances is analyzed theoretically. Section IV summarizes our findings.

## II. METHODOLOGY

### A. Experimental details

The considered multiple QW (MQW) structures consist of 50 periods of GaN/Al$_{0.11}$Ga$_{0.89}$N (11 nm / 3 nm) QWs, grown by plasma-assisted molecular-beam epitaxy along the wurtzite $c$ axis on 4H-SiC substrates with an Al$_{0.11}$Ga$_{0.89}$N (66 nm) buffer layer, as depicted in Fig. 1. The sample referred to as S1 is non-intentionally doped, and S2 is Si-doped with $N_D$=2×10$^{18}$ cm$^{-3}$ in the QW layers. The samples were grown at a substrate temperature of 720°C, under Ga excess and without growth interruptions. More details on the sample growth can be found in Ref. 29. The aluminum content was measured by Rutherford backscattering on 130-nm-thick AlGaN reference layers deposited under the same growth conditions. The structural properties have been analyzed by high-resolution x-ray diffraction (HRXRD) measurements using a Seifert XRD 3003 PTS-HR diffractometer with a beam concentrator prior to the Ge(220) quadruple-bounce monochromator and a Ge(220) double-bounce analyzer in front of the detector.

Figure 2 presents the θ−2θ scan around the (0002) reflection the reciprocal space map around the (-1015) reflection of the GaN/AlGaN MQW. From the inter-satellite distance, we extract a MQW period of 13.5±0.2 nm and 13.6±0.2 nm for S1 and S2, respectively. From the location of the (0002) and (-1015) reflections associated to the AlGaN buffer layer, this layer is about 95% relaxed (i.e. it remains ≈ 5% of compressive strain due to the SiC substrate). Then, by analysis of the angular location of the symmetric and asymmetric MQW reflections, and comparison with simulations (see Fig. 2(a)), the measurements point to an almost complete relaxation of the GaN QWs.

Room temperature (RT) photoluminescence has been measured using pulsed and continuous wave laser operating at 213 and 325 nm wavelengths, respectively. CER spectra has been measured at RT in a capacitor with one semi-transparent electrode made from copper-wire mesh. An alternating voltage (at ~280 Hz) supplied to the capacitor provided the mechanism for band bending modulation inside the sample. White light from a halogen lamp was reflected from the sample surface and directed to a monochromator coupled with a photomultiplier detector. Measurements were performed in a lock-in technique. More details about the CER setup can be found in previous papers.[30]

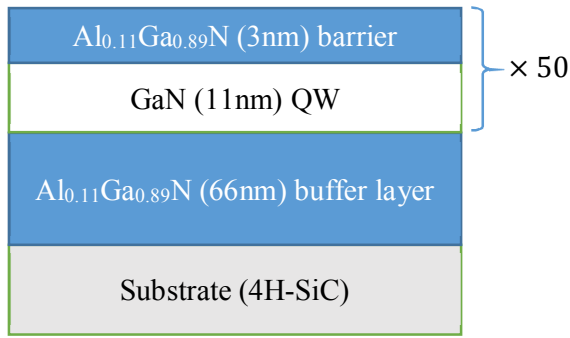

Figure 1. Scheme of the GaN/AlGaN MQWs under study. The sample referred to as S1 is non-intentionally doped, and the sample S2 is Si-doped with $N_D = 2\times10^{18}$ cm$^{-3}$.

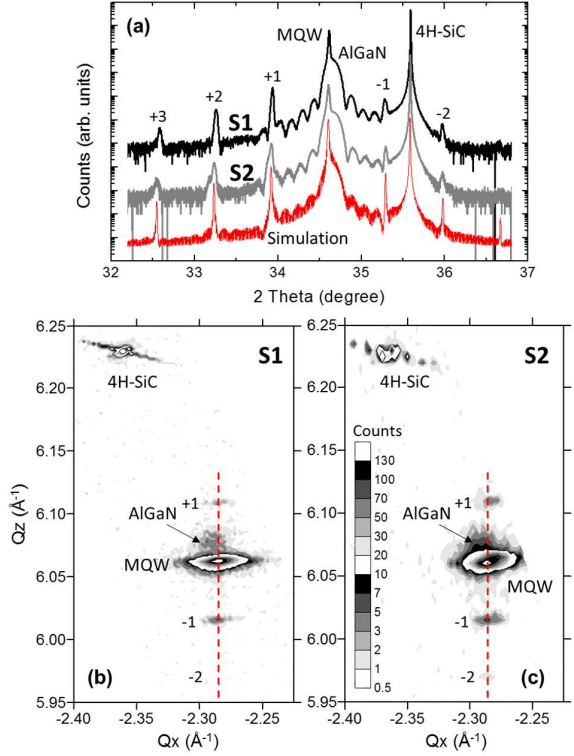

Figure 2. (a) HRXRD θ–2θ scan around the (0002) reflection of the GaN/AlGaN MQWs of samples S1 and S2, together with a simulation generated with the X'pert Epitaxy software from Phillips Analitical. The simulation assumes the GaN layers being fully relaxed and the AlGaN layers fully strained on GaN. (b,c) Reciprocal space maps around the (-1015) reflection of the MQWs of samples (b) S1 and (c) S2. Qx and Qz are the in-plane and out-of-plane reciprocal space vectors.

## B. Theoretical calculations

The electronic band structure of GaN/AlGaN QWs was calculated using the effective-mass approximation. In the valence band, only the heavy-hole band (A band) is considered for simplicity. We take into account the spontaneous and piezoelectric polarization and assume zero potential drop over one MQW period. To account for doping, the Schrödinger equation is solved self-consistently with the Poisson equation, while the Fermi energy $E_F$ is adjusted to preserve charge-neutrality. Calculations are done in the range corresponding to one period of MQW. More detailed description of the calculation method can be found in previous papers.[31,32] The parameters are taken mostly from Ref. 33. From the more recent Ref. 34 by the same authors we take the revised deformation potentials, piezoelectric constants, and the band gap of AlN with bowing parameter, as well as the valence band splitting energies for both binaries. The valence band offset is taken from Ref. 35.

The band structure calculations result in the set of transition energies $E_{ij} = E^e(i) - E^h(j)$ and oscillator strengths (i.e. squared electron–hole overlap integrals) $f_{ij} = \left(\int_{-\infty}^{\infty} h_j(z)\, e_i(z)\, dz\right)^2$, where $E^e(i), E^h(j)$ denote energy and $e_i(z), h_j(z)$ envelope wavefunction of the *i*-th electron and *j*-th hole confined states, respectively. To account for the influence of Fermi energy $E_F$ on the intensity of spectral transition lines, we scale the oscillator strengths by the difference of occupation factors of hole and electron states[36]

$$F_{ij} = f_{ij} \times \left[\phi(E^h(j)) - \phi(E^e(i))\right], \quad (1)$$

where $\phi(E) = \left[1 + \exp\frac{E - E_F}{k_B T}\right]^{-1}$ is the Fermi–Dirac distribution, $T$ is temperature, $k_B$ is Boltzmann constant, and $\phi(E^h(j)) \approx 1$ for *n*-type doping because the hole energy levels are much below the Fermi energy $E^h(j) \ll E_F$.

We simulate the ER spectrum $\Delta R/R$ as a superposition of ER resonance lines,[37] and also calculate the module $\Delta\rho_{ij}$ of each individual resonance line

$$\frac{\Delta R}{R}(E) = \sum_{i,j} Re\left[\frac{F_{ij}\, \Gamma\, e^{i\theta_{ij}}}{(E - E_{ij} + i\Gamma)^n}\right], \quad (2)$$

$$\Delta\rho_{ij}(E) = \frac{F_{ij}\, \Gamma}{\left((E - E_{ij})^2 + \Gamma^2\right)^{n/2}}, \quad (3)$$

where $i = \sqrt{-1}$ and $Re$ denotes the real part. We assume homogeneous broadening parameter $\Gamma = k_B T \approx 25$ meV, and take the exponent $n = 2$ relevant to first derivative Lorentzian excitonic lineshape.[37] In principle, the ER spectrum of QW should be described by the first derivative Gaussian lineshape expressions and Seraphin coefficients.[38] Unfortunately, the coefficients strongly depend on energy due to exciton effects[39], and their accurate calculation is difficult. Therefore, Lorentzian excitonic lineshape is generally used. The parameter $\theta_{ij}$ denotes the unknown phase angle of resonance line, which depends on the measurement set-up and the sample geometry. It is therefore treated as a fitting parameter.

In this paper we neglect the inhomogeneous broadening[28] related to fluctuation of bandgap and QW width because the QW material is a binary compound and a QW width fluctuation of 1-2 ML is negligible given the QW width (11 nm ≈ 40 ML). Possible inhomogeneities result from width fluctuation of the barriers (3 nm ≈ 12 ML), which in polar QWs can affect the transition energies through the electric field distribution,[40] but we estimate that this effect is small compared to RT homogeneous broadening.

## III. RESULTS AND DISCUSSION

### A. Comparison of CER spectra with electronic structure

Figure 3(a) shows the PL and CER spectrum measured at RT for the undoped MQW sample (S1). Due to the fast thermalization of photoexcited carriers from excited to the lowest-energy subbands, the PL spectrum shows the emission from the ground state. The shape of PL peak is significantly different for the two regimes of excitation (pulse and continuous wave excitation), but in both cases the main contribution to emission originates from the recombination between the first electron subband and the first heavy-hole subband. In contrast to PL, optical transitions between ground and excited states are expected in the ER spectrum, because it is an absorption-like experiment. Indeed, the ER signal extends to higher energies than PL. The arrow at 3.65 eV indicates the calculated RT energy gap of $Al_{0.11}Ga_{0.89}N$ fully strained on GaN. In contrast to non-polar QWs, where the distinct transition lines could be easily distinguished in ER,[3-5] here the interpretation of the spectrum is more difficult. According to the principle of reproducing spectrum using a minimal possible number of parameters, four resonances have been used to fit the ER spectrum with Eq. (2). The corresponding moduli of the resonances (Eq. (3)) are shown by dashed lines in Fig. 3(a). The interpretation of these resonances is the main subject of this study.

Figure 3(b)-(d) show the resonance line moduli (Eq. (3)) of all possible interband transitions, grouped by the involved electron level for more clear presentation. The panel (b) shows the transitions from all confined heavy hole levels to the lowest electron level, and panels (c) and (d) to the second and third electron level, respectively. The position of maximum of each curve corresponds to the transition energy, and the height at maximum is proportional to oscillator strength. For each electron level there is a group of consecutive hole levels that contribute with larger oscillator strength than the remaining hole levels, see the moduli for h6-e1, h3-e2, h1-e3 and h5-e3 transition. The transition energies and electron–hole overlaps have been obtained from the calculated electronic structure shown in Fig. 3(e).

The calculated fundamental transition energy agrees with the PL peak within the accuracy of input material parameters and the exciton binging energy. We point out that the ground state transition is about two orders of magnitude weaker than the strongest transitions. This is caused by the very small overlap between e1 and h1 wavefunctions, as shown in Fig. 3(e). It is evident that the first (lowest-energy) fitted ER resonance cannot be attributed to the ground state transition. Moreover, the calculated energy of the ground state transition is below the lowest energy of the fitted resonances. Therefore, the common practice of assigning the ER lowest-energy resonance to the fundamental transition may lead to an overestimation of the Stokes shift.

On the other hand, we can clearly see that there are many confined states and optical transitions between these states are possible in the polar QWs. Some of these transitions strongly overlap with each other, i.e. their energy separation is much less than their widths (FWHM = $2\Gamma \approx 50$ meV). More importantly, several different transitions of comparable strength can contribute to each of the four fitted resonances. We point out that the possibility to get a good fit of the experimental spectrum using four resonances does not mean that there are just four dominant transitions. An indication that there are more transitions involved comes from the fact that the broadening parameters of the fitted resonances are too large to be explained by inhomogeneous broadening of individual transitions. Therefore, each resonance should not be interpreted as a single transition, but rather as a bunch of transitions that are close in energy.

Figure 4 shows the corresponding results for the Si-doped MQW (S2). As in Fig. 3, the CER spectrum has been fitted using four resonances. In Fig. 4(e) we can observe the partial screening of the built-in electric field which leads to a slight increase of the GS electron–hole overlap. Indeed, the module of GS transition (Fig. 4(b)) increased about five times, but it is still much weaker than the module for excited state transitions. Similarly to the undoped MQW case, the lowest-energy fitted resonance cannot be attributed to the GS transition. The built-in electric field is still significant and breaks the square-QW selection rules leading to optical transitions between different hole and electron subbands.

Figure 4(e) shows also the position of the Fermi energy with respect to confined-state energy levels. The influence of Fermi energy and occupation of states on the strength of optical transitions is included in calculations, but the impact is small in this case (the occupation factor of the first electron level is $\phi(E^e(1)) \approx 0.33$, see Eq. (1)). If the doping level were higher enough to bring the e1 level below the Fermi energy, all transitions to e1 would be attenuated. In our case, the doping-induced screening of electric field is not able to restore the GS transition in absorption-like spectra, including CER experiments.

We conclude that the carriers introduced by a doping level of $2\times10^{18}$ cm$^{-3}$ lead only to a slight screening of the built-in electric field. Therefore, doping does not qualitatively change the situation compared to the undoped QW and our previous argumentation holds also for S2.

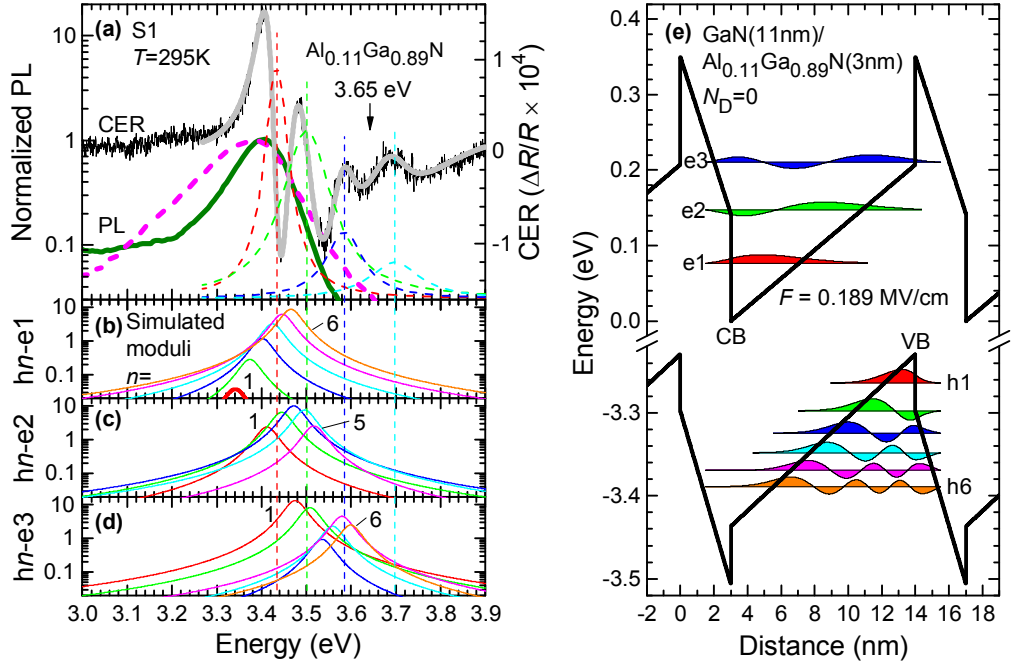

Figure 3. Undoped QW (S1): (a) room temperature PL spectrum measured at 325 nm continuous wave excitation (thick dark green line) and 213 nm pulsed excitation (thick pink dashed line), CER spectrum (solid black line), a fit of CER spectrum (thick gray line), and corresponding moduli of fitted resonances (dashed lines). Vertical lines centered at CER fit moduli peaks are guides to the eye. The arrow indicates the barrier energy gap; (b) simulated moduli of individual transitions from $n$-th heavy hole level to the ground electron level. The ground state transition is highlighted by thick line; (c)-(d) same as in (b) for second and third electron level; (e) electronic structure calculated for one period of MQW: conduction and heavy-hole band profiles shown by solid lines; energy levels and wavefunctions are represented by filled shapes.

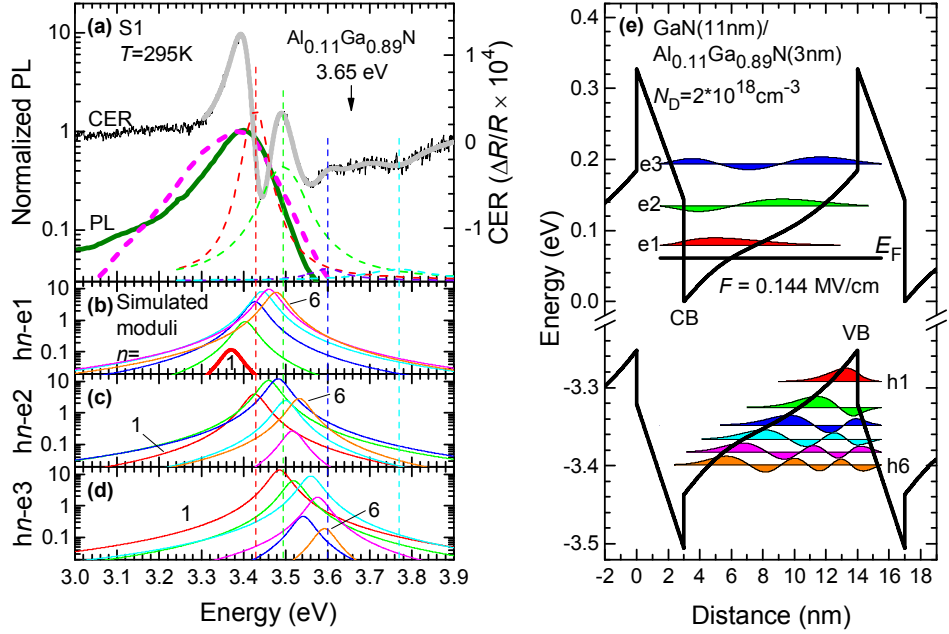

Figure 4. Si-doped QW (S2): (a) room temperature PL spectrum measured at 325 nm continuous wave excitation (thick dark green line) and 213 nm pulsed excitation (thick pink dashed line), CER spectrum (solid black line), a fit of CER spectrum (thick gray line), and corresponding moduli of fitted resonances (dashed lines). Vertical lines centered at CER fit moduli peaks are guides to the eye. The arrow indicates the barrier energy gap; (b) simulated moduli of individual transitions from $n$-th heavy hole level to the ground electron level. The ground state transition is highlighted by thick line; (c)-(d) same as in (b) for second and third electron level; (e) electronic structure calculated for one period of MQW: conduction and heavy-hole band profiles shown by solid lines; energy levels and wavefunctions are represented by filled shapes.

## B. Calculation-based reproduction of the CER spectra

In the following we show that the ER spectrum can be reproduced based on the transition energies and oscillator strengths calculated in the previous section. The spectrum was generated as follows. We set all broadening parameters to $\Gamma = 25$ meV. We selected the 15 strongest transitions. The fitting parameters were the 15 phase angles and a common scaling factor. The total number of fitting parameters was the same as in the case of fitting by 4 resonances in Fig. 3 (it is 15 phase angles and the scaling factor vs. 4 fitting parameters for each of four ER resonances).

Figure 5 shows the comparison of the experimental CER spectrum and our calculation using 15 transitions. The overall accuracy of the reproduction is very good, considering that all the energies, relative strengths, and broadening widths have been fixed to theoretical values. Nevertheless, there are some deviations above 3.55 eV which might be caused by transitions to higher states, i.e. so called above barrier transitions, which are not taken into account in this simulation.

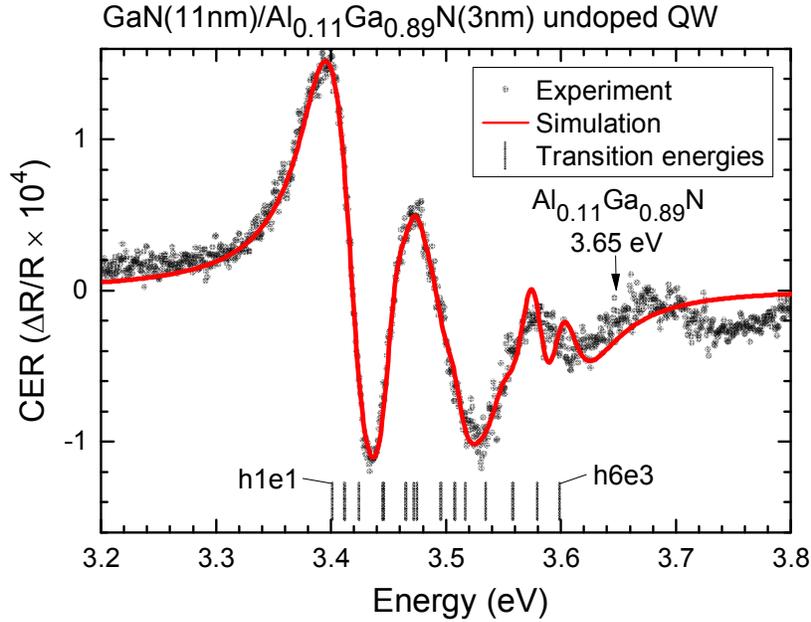

Figure 5. Reproduction of the CER spectrum using the theoretically calculated transition energies and oscillator strengths, assuming homogeneous broadening (Γ = 25 meV). Lines at the bottom mark the energies of the 15 strongest interband transitions taken into account. The phases and a common scaling factor have been obtained by a fitting procedure.

## C. Examples of merging ER resonances

Experimentally it is not possible to observe the intermediate steps of the merging of ER resonances separated by small energies since an independent control of transition energies, oscillator strengths, and broadenings in GaN/AlGaN QW samples is very difficult to obtain. To illustrate this process, in this section we generate the ER spectra composed of a few resonances. Figure 6 shows spectra constructed from four resonances of equal oscillator strength and broadening parameter Γ. In general, the oscillator strength varies from transition to transition due to their different electron-hole overlaps and their different sensitivity to electro-modulation in ER measurements. However, to illustrate the problem of superposition/merging ER resonances, we have neglected these differences. The transition energies in our simulations are $E_i = 3.4 \text{ eV} + i \, \Delta E, i = 0, \ldots, 3$, and we compare different values of energy separation, $\Delta E$. The appearance of resulting spectrum depends on the phase of the individual resonances, but as an example we selected $\theta_i = 0, \frac{\pi}{2}, \pi, \frac{3\pi}{2}$. For $\Delta E \geq 2\Gamma$, it is clear that the merged spectrum in Fig. 6(a) consists of several resonances, but for $\Delta E = \Gamma$, the spectrum in Fig. 6(c) looks like a single spectral line, though unusually shaped. This means that considering the overlapping of ER resonances is particularly important for proper analysis of ER spectra when the energy separation between the transition levels is comparable or smaller than the broadening of ER resonance, which is assumed in this case to be $k_B T \approx 25$ meV.

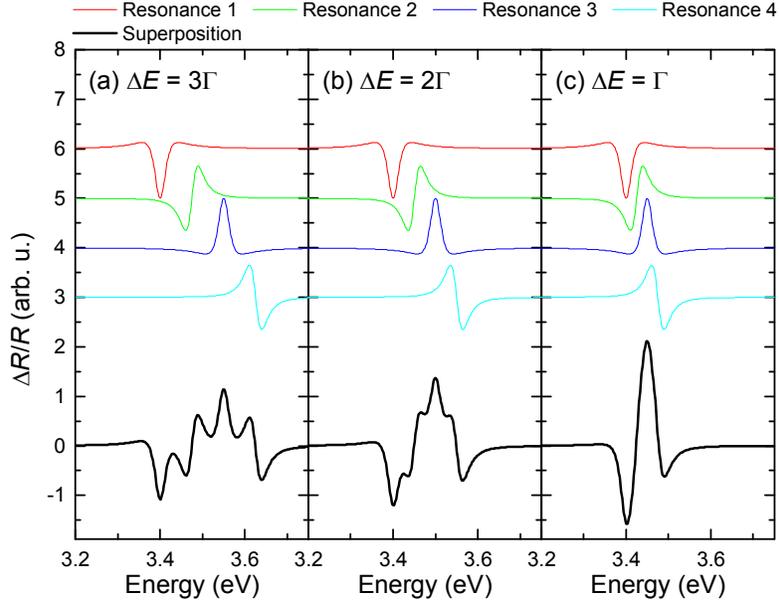

Figure 6. ER spectrum composed from four resonances of width $FWHM = 2\Gamma = 50$ meV which are separated by different $\Delta E$: (a) $\Delta E = 0.075$ eV $= 3\Gamma$, (b) $\Delta E = 0.05$ eV $= 2\Gamma$, (c) $\Delta E = 0.025$ eV $= \Gamma$.

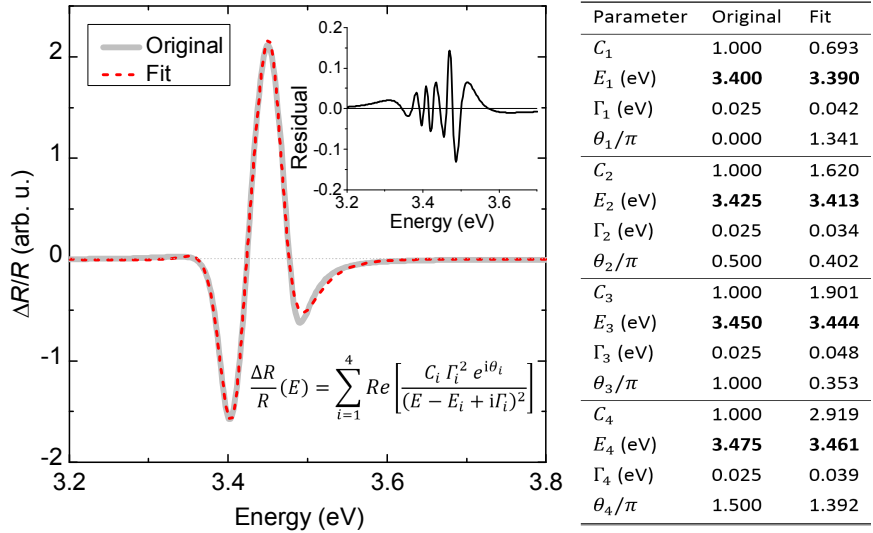

Figure 7. Fitting of spectrum consisting of four partially overlapping resonances and comparison of parameters used to generate the spectrum with those obtained from the fitting procedure.

While in Fig. 6(a) the transition energies can be resolved even by naked-eye, for smaller energy separation one must rely on a fitting procedure. It is interesting to check if the fitting procedure can reproduce the original values of parameters. Figure 7 shows the original spectrum and the result of fitting by 4 resonances, where each has 4 fitting parameters (strength $C_i$, energy $E_i$, broadening $\Gamma_i$, phase $\theta_i$, $i = 1, ..., 4$). The reproduced graph looks similar, but in the table we can see that the fitting parameters differ from the original ones. Although the graphical comparison suggests good accuracy of the fit (the residual of the fit is around 5% of the signal, which is often less than the signal to noise ratio) the uncertainty in fitted energies reaches half of the energy separation between transitions. It means that the fitted energies are quite arbitrarily positioned in the considered spectral range. The next signature of inaccurate fitting is that the broadening parameters are overestimated. Moreover, the fitted amplitudes and phases are not correlated with the original values. In summary, it is important to be aware that in the case of strong overlap the fitted parameters are not directly related to the actual transitions that generate the spectrum. In such a situation, the ER spectrum can be reproduced by a single resonance which energy corresponds to an average energy of the four optical transitions involved.

## IV. SUMMARY

We have analyzed CER and PL spectra measured at room temperature for polar GaN/AlGaN QWs, both undoped and Si-doped. The band structure calculations indicated very weak electron–hole wavefunction overlap for the GS transition in these 11 nm wide QWs, which has major implications for interpretation of CER spectra. The fundamental transition was too weak to be observed in the CER spectrum. This effect gives rise to an intrinsic Stokes shift between emission and absorption-like spectra. The built-in electric field breaks the square-QW selection rules giving rise to a number of optical transitions between different hole and electron subbands, which are relatively close in energy. Based on their calculated energies and strengths, we were able to reconstruct the CER spectrum. In standard approach, the CER spectrum is modelled by considering the minimum number of resonance line-shapes that provide a good fit to the experiment (in our case four). However, we have found that several transitions can contribute to each of the fitted resonances. This is evidenced by their large FWHM of about 100-150 meV, much larger than the 2kT≈50 meV linewidth expected for individual transitions with weak inhomogeneous broadening. In this case, proper interpretation of the fitted resonances requires analysis of their broadening parameters and comparison with calculated interband transition energies in the corresponding energy range.

**Acknowledgments**

This work was performed within the grant of the National Science Centre, Poland (grant no. 2016/21/B/ST7/01274).